# Observation of topological edge modes in bianisotropic metamaterials

Alexey P. Slobozhanyuk[1,4]†, Alexander B. Khanikaev[2,3]†, Dmitry S. Filonov[4],

Daria A. Smirnova[1], Andrey E. Miroshnichenko[1], and Yuri S. Kivshar[1,4]*

[1]Nonlinear Physics Center, Australian National University, Canberra ACT 0200, Australia.

[2]Department of Physics, Queens College of City University of New York, NY 11367, USA.

[3]Department of Physics, Graduate Center of City University of New York, NY 10016, USA.

[4]ITMO University, St. Petersburg 197101, Russia

*Correspondence to: ysk@internode.on.net

†These authors contributed equally to this work

**Abstract**

Existence of robust edge modes at interfaces of topologically dissimilar systems is one of the most fascinating manifestations of a novel nontrivial state of matter, *topological insulators*. Such electronic states were originally predicted and discovered in condensed matter physics, but they find their counterparts in other fields of physics, including the physics of classical waves and electromagnetics. Here, we present the first experimental realization of a topological insulator for electromagnetic waves based on engineered bianisotropic metamaterials. By employing the near-field scanning technique, we demonstrate experimentally the topologically robust propagation of electromagnetic waves around sharp corners without back reflection.

Topologically nontrivial states of light[1] represent a "Holy Grail" for optical applications usually suffering from undesirable backscattering and interference effects, thus dramatically limiting the bandwidth and performance of many photonic devices. Recently there were numerous theoretical predictions of topologically nontrivial states in different photonic systems ranging from microwave to optical frequencies[2-10], but experimental demonstrations are very limited. In particular, the topological protection was shown experimentally for gyromagnetic photonic crystals[12,13], arrays of coupled ring resonators[14], chiral optical fibers[15], and microwave waveguides[16]. In all such systems, different approaches were suggested to emulate an effective magnetic field, in a direct analogy with the familiar Quantum Hall Effect (QHE). However, another class of electromagnetic systems for which topological nontrivial states emulating Quantum Spin Hall Effect (QSHE)[17-22] and robust helical edge modes have been predicted represents a bianisotropic metamaterial[9,10] - a special class of synthetic periodic optical media with engineered magneto-electric response[23-26]. Here we suggest and demonstrate the first experimental realization of topologically nontrivial metamaterials which emulate directly the spin-orbit coupling in solids and exhibit the desirable property of topologically robust edge transport enabling guiding waves through sharp corners.

The topological electromagnetic states realized in our system are based on two key ingredients: (i) "spin-degeneracy" enabling emulation of the electron spin, and (ii) bianisotropic gauge fields mimicking the spin-orbital coupling for electrons in solids. We consider a metacrystal with a square lattice of metamolecules with the shape of a ring and two symmetric slits and a wire placed in the middle. The schematic of the single metamolecule and the periodic arrangement of the metacrystal are shown in Fig. 1. The metamolecule of this geometry possesses two low-frequency magnetic and electric dipolar resonances with their field profiles shown in Fig. 1b. The magnetic resonance is formed by the antisymmetric charge distribution in the wires of the split-ring and with the inactive central wire (an upper panel in Fig. 1b). The magnetic dipole moment of this mode is oriented in the direction orthogonal to the plane of the metamolecule and originates from the currents counter-propagating in the wires. It is also easy to see that the net electric dipole moment of this mode vanishes due to cancellation of dipole moments of the individual antennas of the split-ring. In contrast, the electric dipolar resonance originates from the dipole moment of the central wire dressed up by the interaction with the wires of the split-ring.

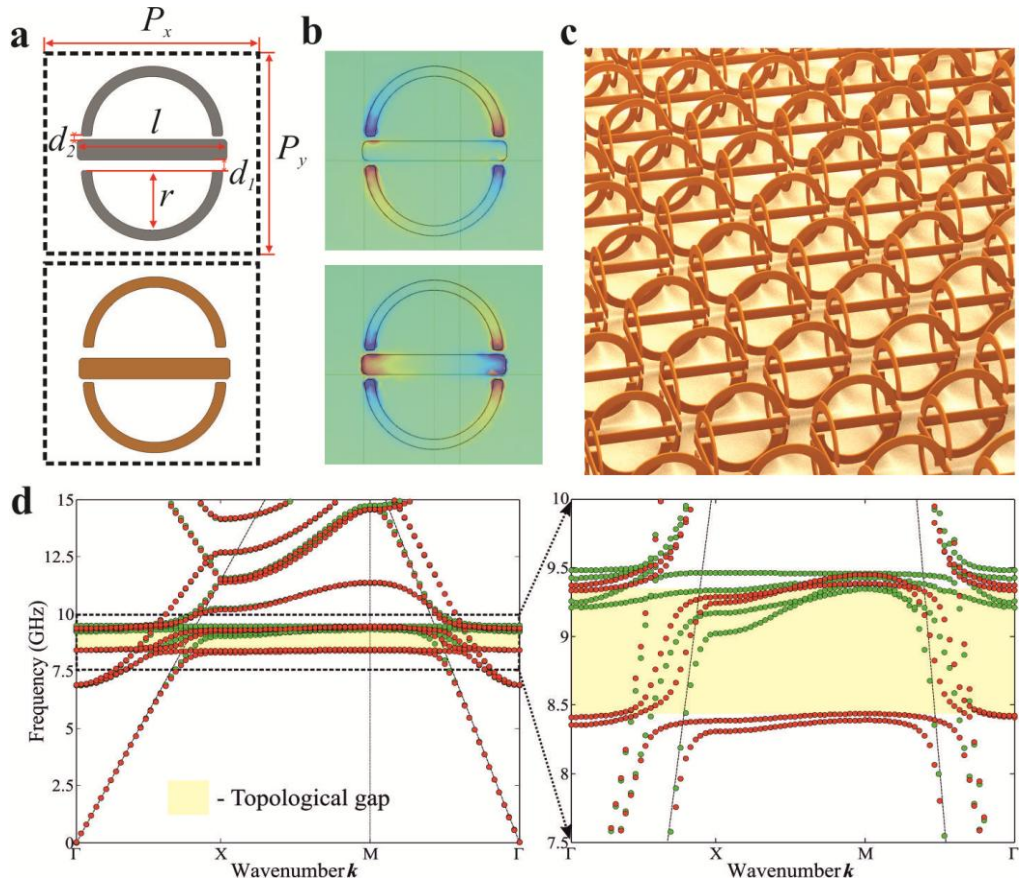

**Figure 1 | Structure and eigenmodes of the bianisotropic metamolecules and a metacrystal. a,** Geometry of the metamolecules with bianisotropic response controllable by the asymmetry of $d_1$ and $d_2$ gaps. The radius ($r$), length of the wire ($l$), and periods ($P_x$, $P_y$) are equal to 2.85 mm, 0.5 mm, 7.5 mm, 13 mm and 12 mm respectively. Metal width for the central wire and split ring wires is 1 mm and 0.5 mm, respectively. **b,** Dipolar magnetic (top subplot) and dipolar electric (bottom panel) eigenmodes of the symmetric metamolecule ( $d_1 = d_2 = 0.325$ mm). **c,** Perspective view of the two-dimensional metacrystal formed by periodic stacking of the metamolecules. **d,** Photonic band structure of the metacrystal with ($d_1 = 0.5$ mm $\neq d_2 = 0.15$ mm) and without ($d_1 = d_2 = 0.325$ mm) bianisotropy shown by red and green markers, respectively. The right subplot shows enlarged region near the topological band gap. The yellow shaded area illustrates the spectral width of the gap. The substrate with the permittivity equal to $\epsilon = 3.38$ is taken into account. Light line position is indicated by thin black dashed lines.

This resonance appears to be shifted significantly towards the subwavelength range thanks to the hybridization with the split-ring by inducing symmetric currents distribution in the wires. The latter feature allows us to satisfy the spin-degeneracy condition as the geometrical parameters of the metamolecule can be tuned to ensure that the electric and magnetic dipolar resonances occur at the same frequency.

As the next step, we form a metacrystal (see Fig. 1c) by arranging the metamolecules into a square two-dimensional array with only one layer in vertical direction. The open geometry considered here represents a special interest due to possible leakage of topological modes caused by the out-of-plane scattering into the radiative continuum. The square lattice of the metacrystal with $C_{4v}$ in-plane and z-inversion symmetries possesses quadratic degeneracies at high symmetry Γ and M points in the Brillouin zone[4,12,27], which is confirmed for both electric and magnetic dipolar modes of the metacrystal by first principle finite element method simulations (see Methods for details), and are shown in Fig. 1d (green dotted line). Here we are interested in guided waves exponentially confined to the structure in vertical (z-) direction and whose bands are located below the light cone (dashed straight lines in Fig. 1d), and, therefore, in what follows we focus on the quadratic degeneracies taking place at M points. Note, that to ensure the condition of spin-degeneracy at M-point, e.g. the two doubly-degenerate quadratic bands stemming from magnetic and electric resonances appear at the same frequency, we fine-tuned the geometry of the metamolecules to compensate for the effects of slightly different interaction between magnetic and electric dipoles in the array. The resultant band structure (shown in Fig. 1d) confirms the presence of overlaid electric-like and magnetic-like collective modes with quadratic degeneracy at the M point.

The presence of the quadratic degeneracies is crucial for topological protection in the metamaterial under study; to illustrate this, here we first apply the analytic effective Hamiltonian description[3,9,27]. The band structure of the metacrystal in the proximity of the M-points can be described by the effective quadratic Hamiltonian $\widehat{\mathcal{H}}_0 = \{\widehat{\mathcal{H}}_e\ 0; 0\ \widehat{\mathcal{H}}_m\}$, acting on a four component wavefunction $\boldsymbol{\psi}_{em} = \{e_x, e_y, m_x, m_y\}$, with its components being in-plane electric and magnetic dipole moments, and subscripts $e$ and $m$ indicating the electric and magnetic blocks, respectively, which are described by the expression[27]:

$$\widehat{\mathcal{H}}_{e/m} = \lambda_{e/m}\big[\beta_{e/m}(\delta k_x^2 - \delta k_y^2)\hat{\sigma}_x + 2\delta k_x \delta k_y \hat{\sigma}_y + \gamma_{e/m}|\delta\boldsymbol{k}|^2\big]. \qquad (1)$$

Here $\hat{\sigma}_i$ are the Pauli matrices, $\delta\boldsymbol{k}$ is the deviation of the Bloch wavevector from the M-point, and $\lambda_i, \beta_i$, and $\gamma_i$ are the effective parameters of the model, which in general differ for electric and magnetic modes. However, in our design the dispersion of the electric and magnetic modes has been tuned to match near the M-point so that these parameters are, to a good approximation, are equal for both types of modes. This degeneracy is crucially important as it allows choosing a new set eigenmodes as any linear combination of electric and magnetic components. One of the key ingredients for engineering the topological state in the metacrystal represents the ability to emulate the spin degree of freedom of electrons in QSHE, which must be odd under time reversal (TR) operation. By noticing that electric and magnetic moments transform differently under TR $Te_i = e_i$ and $Tm_i = -m_i$, it was shown that an appropriate choice of basis is given by the wavefunction $\boldsymbol{\psi}_\pm = \{e_x + m_x, e_y + m_y, e_x - m_x, e_y - m_y\} = \{\psi_x^+, \psi_y^+, \psi_x^-, \psi_y^-\}$, where $\psi_i^+$ and $\psi_i^-$ are the desirable pseudo-spin components which are time-reversal partners $T\psi_i^\pm = \psi_i^\mp$ of each other[9].

Next, to endow the metacrystal with topological properties, we introduce an effective gauge field emulating the spin-orbital coupling, which is achieved by introducing bianisotropy mixing magnetic and electric degrees of freedom of the system[9]. The desirable form of bianisotropy should couple x-(y-) magnetic dipole moment with y-(x-) electric dipole moment of the metamolecule, and is realized by breaking its z-inversion symmetry[23]. Figure 1d (red dotted line) shows the band structure calculated by the FEM for the case of distorted metamolecules, where the central wire was displaced vertically between the split ring wires (so that gaps $d_1$ and $d_2$ are no longer equal), and reveals a gap that is complete below the light cone. In the context of the effective Hamiltonian description the effect of the bianisotropy induced by this symmetry reduction can be described by an spin-orbital-like potential $V^{SO} = \{\zeta\hat{\sigma}_z, 0; 0, -\zeta\hat{\sigma}_z\}$ [9], which results in opening of the photonic band gap and transition from topologically trivial to topological nontrivial state (see Supplementary Information for details).

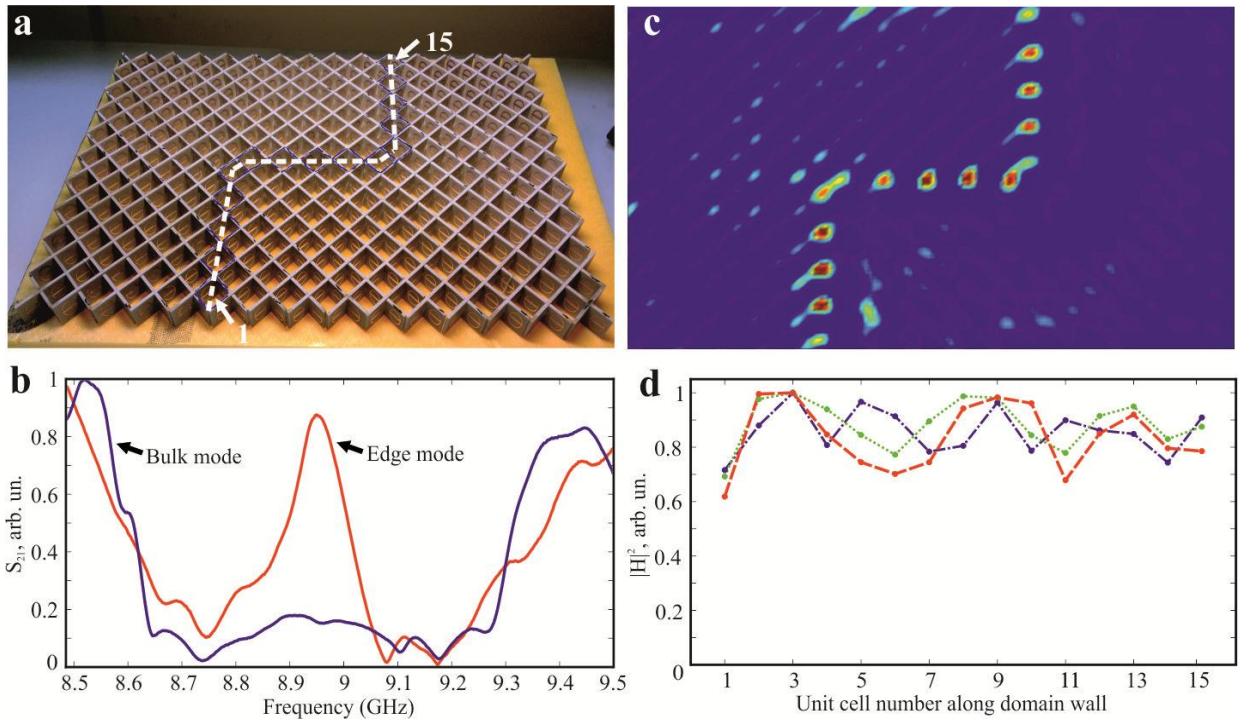

**Figure 2 | Experimental system and the observation of topological edge modes**. **a,** Photograph of the fabricated metacrystal with the location of the double-bend domain wall indicated by the white dashed line. **b,** Transmission spectra of the metacrystal away from the domain wall (blue line) and along the domain wall (red line). **c,** Two-dimensional map of the magnetic field intensity indicating reflection-less propagation along the domain wall. **d,** Normalized magnetic field intensity along the domain wall measured for different frequencies (8.9 GHz - blue line; 8.95 GHz - green line; 8.987 GHz - red line) within the topological band gap.

Appearance of topologically robust helical edge modes insensitive to local structural imperfections and avoiding backscattering represents a hallmark of topological photonic order emulating QSHE[8,9,28]. In the case of bianisotropic metacrystals[9], the edge states have been predicted to occur at the "domain walls" representing boundaries between regions of the metacrystal with the reversed bianisotropy. The presence of such modes was confirmed by calculating the photonic band structure in the effective Hamiltonian model, as described in the Supplementary Information. From the bulk-boundary correspondence, for such a domain wall, one expects two edge modes for every spin, in accordance with the change in the spin-Chern number between the domains (ref. 9 and Supplementary Information). It is worth mentioning that one could also expect to observe the edge modes on the external boundary of the metacrystal with the air, however, in contrast to electronic systems, the external domain does not possesses a bandgap in the spectrum and is filled with the electromagnetic continuum. This makes such modes either leaky or unstable to any perturbation which couples them to the radiative continuum of the free space, thus breaking the topological protection.

To verify these theoretical predictions and confirm topological robustness of the helical edge states, a metacrystal was fabricated by printing an array of metamolecules on the a 1 mm-thick dielectric board (Arlon 25N) with the dielectric permittivity $\epsilon = 3.38$. The board was cut into linear segments which were stacked to form a square lattice, as shown in Fig. 2a. To test the topological protection, which leads to the ability of the edge modes to propagate around sharp bends without back scattering[9], the zigzag-shaped domain wall was created by deliberate distribution of metamolecules with the central wires shifted vertically up and down across the crystal, as indicated by the dashed white line in Fig.2a.

First, the presence of the topological band gap was verified by exciting bulk modes with the dipole source placed away from the domain wall. Figure 2b clearly reveals the gap spanning frequency range from 8.6 GHz to 9.3GHz. Next, the presence of the topological edge mode was tested by placing the dipole at the domain wall and the transmission spectrum was measured. As expected, the enhanced transmission within the band gap region occurred due to the excitation of the edge mode, as shown in Fig. 2b. Note that the transmission was especially high near the gap center and gradually dropped towards the edges of the bands.

To confirm that the transmission indeed occurs due to the edge mode localized to the domain wall, the near-field map of magnetic field intensity was measured across the entire sample with the use of the probe (magnetic loop antenna) mounted on the two-dimensional near-field scanning stage. The map shown in Fig 2c clearly shows that the mode excited at the frequency of $f = 8.987$ GHz is indeed guided by the domain wall. In agreement with the theoretical predictions, the mode appears to be strongly localized to the wall. In addition, quite surprisingly we have not noticed any apparent effects of leakage neither to the modes above the light cone nor to the radiative continuum. Moreover, we observed that regardless of the open character of the system, the edge modes do not scatter to these modes even when they encounter sharp bends of the domain wall, and the wave flawlessly transfers between orthogonal segments of the wall. To ensure the robustness across the entire frequency range of the topological bandgap, the same measurements have been conducted for multiple frequencies. As expected for the frequencies close to the spectral edges of the bulk modes, the edge states become poorly localized and are hardly distinguishable from the bulk modes. However, as can be seen from Fig. 2d, the edge states with frequencies sufficiently apart from the bulk spectrum remains well localized and exhibits similar robustness against sharp bends of the domain wall at all frequencies. It is also important to mention that the drop in the transmission observed in Fig. 2b away from the center of the topological band gap lacks back-reflection along the domain wall (observed in Fig. 2d), which indicates that the low transmission near the edges is associated with the insertion loss. This can be explained as the consequence of increasingly poor overlap of the field produced by the dipole antenna with the field profile of the edge mode which spread over the bulk as we approach the band edges. Also note that in Fig. 2d some variations in the field intensities occur due to the local interference effects, which, however, do not lead to backscattering, but rather assist total transmission through the zigzag, as is evidenced by the rebound of the intensity away from the $90^o$-bends of the wall. Finally, it's important to mention that according to our measurements the finite lateral size of the sample did not play any detrimental role and no any noticeable backscattering of the edge modes occurred at the metacrystal boundaries, where the edge modes were efficiently radiated into the propagating continuum of the free space.

Topological robustness of electromagnetic helical edge-states in an open metamaterial system demonstrated here for the first time can be of significant interest both from fundamental and applied perspectives. First, it confirms the possibility to emulate exotic quantum states of solids in open metamaterial structures thus allowing direct near-field mapping of the local amplitudes and phases of topological waves. This enables the study of their topological characteristic and peculiarities in the real coordinate space as well as in the reciprocal space by performing Furrier transformation of the measured field maps. Second, the possibility to guide electromagnetic energy around arbitrarily shaped pathways avoiding undesirable back-reflection from sharp bends and without leakage into the radiative continuum brings the versatility of control of the electromagnetic energy flows in engineered photonic systems to an unprecedented level. It is only a matter of time when the full potential of topological photonic states will be exploited to study novel fundamental electromagnetic phenomena and in applied systems and devices with their physics and functionality solely relaying on the presence of the photonic topological order.

## Methods

**Simulations.** All numerical results for photonic band structures were obtained by performing finite-element-method calculations in COMSOL Multiphysics (Radio Frequency module). We have assumed perfectly conducting elements (split ring and wire) on the dielectric substrate. The periodic boundary conditions were imposed in $x$ and $y$ directions to form an infinite square lattice (shown in Fig. 1c). The perfect matched layers were added to the domain in order to prevent back reflections from $z$ direction. The mesh was optimized in order to reach the convergence.

**Experiments.** The metamolecules were fabricated on the dielectric board (Arlon 25N) using chemical etching technique. Subsequently, the board was cut into linear segments which were staked to form a square lattice. Special mask made from styrofoam material with the dielectric permittivity of 1 was used to control the periodicity of the segments separation. All the measurements were performed in the anechoic chamber. We utilized subwavelength dipole as a source, which was connected to the transmitting port of a vector network analyser (Agilent E8362C). For measurement of the transmission spectra shown in Fig. 2b a similar dipole antenna was used as a receiver. To perform the near field measurements we used an automatic mechanical near-field scanning setup and a magnetic field probe connected to the receiving port of the analyser (Fig.2c). The probe was oriented normally with respect to the interface of the structure. The near field was scanned at the 1 mm distance from the back interface of the metacrystal to avoid a direct contact between the probe and the sample.

**Acknowledgements**

This work was supported by the Australian Research Council, USA Air Force Research Laboratory, the Government of the Russian Federation (Grant 074-U01), and the Dynasty Foundation (Russia). APS acknowledges a support of the SPIE scholarship. The authors are grateful to Dr. Alexander Poddubny and Dr. Ilya Shadrivov for useful discussions and suggestions.